\documentclass[conference]{IEEEtran}
\IEEEoverridecommandlockouts
\usepackage{cite}
\usepackage{amsmath,amssymb,amsfonts}
\usepackage{algorithmic}
\usepackage{graphicx}
\usepackage{textcomp}
\usepackage{xcolor}
\def\BibTeX{{\rm B\kern-.05em{\sc i\kern-.025em b}\kern-.08em
    T\kern-.1667em\lower.7ex\hbox{E}\kern-.125emX}}
\begin{document}

\title{Ask What Your Country Can Do For You: Towards a Public Red Teaming Model\\

\thanks{Presented at the Conference on Applied Machine Learning in Information Security (CAMLIS) RED Workshop, 2025.}
}
\author{\IEEEauthorblockN{Wm. Matthew Kennedy}
\IEEEauthorblockA{\textit{Oxford Internet Institute} \\
\textit{University of Oxford}\\
Oxford, UK\\
matt.kennedy@oii.ox.ac.uk}
\and
\IEEEauthorblockN{Cigdem Patlak}
\IEEEauthorblockA{\textit{Independent} \\
Irvine, USA \\
}
\and
\IEEEauthorblockN{Jayraj Dave}
\IEEEauthorblockA{\textit{Independent} \\
Dallas, USA \\
}
\and
\IEEEauthorblockN{Blake Chambers}
\IEEEauthorblockA{\textit{Independent} \\
Boston, USA \\
blakejwc@alum.mit.edu}
\and
\IEEEauthorblockN{Aauysh Dhanotiya}
\IEEEauthorblockA{\textit{Amazon} \\
Seattle, USA \\
adhanoti@alumni.cmu.edu}
\and
\IEEEauthorblockN{Darshini Ramiah}
\IEEEauthorblockA{\textit{Infocomm Media Development Authority} \\
Singapore\\
Darshini\_Ramiah@imda.gov.sg}
\and
\IEEEauthorblockN{Reva Schwartz}
\IEEEauthorblockA{\textit{Civitaas Insights LLC} \\
Washington, D.C., USA \\
reva@civitaas.com}
\and
\IEEEauthorblockN{Jack Hagen}
\IEEEauthorblockA{\textit{Department of Computer Science} \\
\textit{University of Wisconsin - Eau Claire}\\
Eau Claire, USA \\
hagenjj4111@uwec.edu}
\and
\IEEEauthorblockN{Akash Kundu}
\IEEEauthorblockA{\textit{Humane Intelligence} \\
akash@humane-intelligence.org}
\and
\IEEEauthorblockN{Mouni Pendharkar}
\IEEEauthorblockA{\textit{Independent} \\
San Francisco, USA}
\and
\IEEEauthorblockN{Liam Baisley}
\IEEEauthorblockA{\textit{Carnegie Mellon University} \\
Pittsburgh, USA \\
}
\and
\IEEEauthorblockN{Rumman Chowdhury}
\IEEEauthorblockA{\textit{Humane Intelligence and Harvard University} \\
\textit{Berkman Klein Center for Internet and Society}\\
New York, USA \\
rumman@humane-intelligence.org}
\and
\IEEEauthorblockN{Theodora Skeadas}
\IEEEauthorblockA{\textit{Humane Intelligence} \\
Boston, MA\\
theodora@humane-intelligence.org
}
}

\maketitle



\section{Overview}
AI systems have the potential to produce both benefits and harms, but without rigorous and ongoing adversarial evaluation, AI actors will struggle to assess the breadth and magnitude of the AI risk surface. Researchers from the field of systems design have developed several effective sociotechnical AI evaluation and red teaming techniques targeting bias, hate speech, mis/disinformation, and other documented harm classes. However, as increasingly sophisticated AI systems are released into high-stakes sectors (such as education, healthcare, and intelligence-gathering), our current evaluation and monitoring methods are proving less and less capable of delivering effective oversight [2, 10]. 

This problem is compounded by the perception that AI labs are solely responsible for delivering AI safety and security – in fact, this duty is shared by AI application developers and public entities alike [13]. Yet, those able to perform expert AI evaluations or red teaming exercises at scale are primarily employed by AI labs themselves. In turn, AI application developers may be unable or unwilling to pursue state-of-the-art evaluations or red teaming programs, trusting instead that the AI models they have built upon are sufficiently safe and secure. Moreover, public entities have not typically thought to contribute to this function directly through technical interactions during the AI design, development, or deployment stages.
In order to actually deliver responsible AI and to ensure AI’s harms are fully understood and its security vulnerabilities mitigated, pioneering new approaches to close this “responsibility gap” are now more urgent than ever [10]. In this paper, we propose one such approach, the cooperative public AI red-teaming exercise, and discuss early results of its prior pilot implementations. This approach is intertwined with CAMLIS itself: the first in-person public demonstrator exercise was held in conjunction with CAMLIS 2024. We review the operational design and results of this exercise, the prior National Institute of Standards and Technology (NIST)’s Assessing the Risks and Impacts of AI (ARIA) pilot exercise, and another similar exercise conducted with the Singapore Infocomm Media Development Authority (IMDA). Ultimately, we argue that this approach is both capable of delivering meaningful results and is also scalable to many AI developing jurisdictions.

\section{Theory and Design}

AI red teaming is a dynamic new discipline within the broad field of Testing, Evaluation, Validation and Verification (TEVV) [6]. It draws upon traditional adversarial cyber-security techniques to develop an understanding of safety and security vulnerabilities that are unique to AI systems [1]. Although AI red teaming lacks a coherent theoretical foundation, existing practices suggest that its primary objective is to achieve a comprehensive view of the entire “harm surface” of a given AI model or system, a principle often referred to as coverage. To organize attacks and systematize findings, AI red teamers employ sophisticated taxonomies of technical and sociotechnical risks, often formalized as specific attack classes or harm or risk typologies [4]. Among the best known are the MITRE ATLAS and the U.S. National Institute of Standards and Technology (NIST) AI RMF (100-1) [8] or the NIST Generative AI (GAI) Profile (600-1) [9]. 

Of course, red-teaming is not a silver bullet [3]. Several compelling critiques of prevailing red teaming practices have appeared in the discourse. Our exercise attempted to address two of these: 1) that in-house red teaming is inadequate for achieving coverage, as it is either not sufficiently blinded, favors developer-created harm classifications at the expense of classifications that users might articulate, or is carried out by teams that are insufficiently diverse, and 2) that popular risk or harm typologies are too general to be readily operationalizable for most engagements, given the unique norms, values, and discourses of the intended deployment context [5]. It is for these reasons that many AI ethics, sociotechnical safety, metrology, and product professionals have urged AI developers to incorporate public third-party evaluations and red teaming in their broader TEVV function [11, 7]. Despite the broad support for large-scale third-party evaluation and red teaming programs, few attempts have been made to operationalize this new methodology.

\section{Cases and Discussion}
Our presentation discusses the experimental design and findings of three of the largest public third-party AI red teaming exercises on aligned production models yet organized. First, we review the outcomes and lessons of the NIST Assessing the Risks of AI (ARIA) pilot exercise, held virtually in the fall of 2024. Second, we discuss the subsequent in-person red teaming exercise facilitated by an undisclosed U.S. government partner and held alongside CAMLIS 2024. Finally, we discuss the recently concluded IMDA Singapore AI Safety Red Teaming Challenge on multilingual and multicultural bias across nine different countries and languages. 

\subsection{Case 1: NIST ARIA National Red Teaming Pilot Exercise}\label{AA}
In late 2024, Humane Intelligence, a nonprofit whose mission is to develop an AI evaluation community of practice, partnered with a prominent national metrology institution to conduct the first AI red teaming event open to all residents of that institution’s jurisdiction. The operation consisted of a scenario-based capture-the-flag (CTF) exercise. To ensure broad public participation and to recruit a diverse  multidisciplinary red team, enrollment criteria were as permissive as possible: anyone resident in the United States and eighteen years of age or older could enroll in the virtual qualifier. Recruiting broadly also supported the need to expand AI governance into the public sphere. We specifically invited applications from individuals with diverse expertise, including, but not limited to, AI researchers and practitioners, cybersecurity professionals, data scientists, ethicists and legal professionals, software engineers, and policymakers and regulators. 457 participants were enrolled in the event.

The operation was delivered in partnership with the NIST ARIA pilot exercise during September - October 2024. During this exercise, red teaming participants sought to stress test model guardrails and safety mechanisms to produce as many violative outcomes as possible from three LLMs. Models were red teamed in the context of three test scenarios defined by ARIA: inducing a meal planning system to give harmful food-related advice, inducing a travel advisory system to make factual errors or nonsensical travel plans, and inducing a film and TV resource system to reveal spoilers. We discuss the results of this exercise as well as the strengths and limitations of operational design with a specific emphasis on assessing the benefits provided by marshaling a large (if also contingent), diverse, and multidisciplinary red team.

\subsection{Case 2: Public Red-Teaming Demonstrator Exercise at CAMLIS 2024}\label{AA}

Subsequently, Humane Intelligence and an undisclosed U.S. government partner organized an in-person red teaming exercise, which was held during the CAMLIS cybersecurity conference on October 24-26, 2024. The two objectives of this in-person event were: (1) to assess the vulnerabilities of each AI model and model provider’s AI security posture; and (2) to use and assess the utility of NIST AI 600-1 as a framework for evaluating GAI risks and suggested mitigations. The operation consisted of a series of in-person red teaming scenarios, each targeting different office productivity software that employed GenAI models. Model owners applied to be part of this exercise, and each application was evaluated according to specific criteria.

Targets comprised of models or applications submitted for testing by four AI model owners. Models possessed various capabilities including text-to-video generation and reasoning over documents via RAG. Red teamers were selected from the 30 highest scoring participants in the NIST ARIA pilot exercise, which yielded a team from a broad spectrum of backgrounds, cultures, and disciplines. Divided into two teams, red teamers were instructed, firstly, to use a dedicated testing UI to induce submitted models to produce violative outputs or behaviors described by six risk categories derived from a leading AI risk typology. Second, they were instructed to analyze the utility of that risk typology as a red teaming framework for each operation and risk category. We discuss the scientifically significant results of this exercise, as well as the relative strengths and weaknesses of an intensive in-person operation versus an operation carried out over a longer period of time in a nationally distributed manner.

\subsection{Case 3: IMDA Singapore AI Safety Red Teaming Challenge}\label{AA}
In the final months of 2024, Humane Intelligence partnered with the Singapore Infocomm Media Development Authority (IMDA) to hold the world's first-ever multilingual and multicultural AI safety red teaming exercise focused on Asia-Pacific. Through this exercise, which was carried out both in person and virtually, partner institutes developed a methodology for evaluating LLMs for context-specific harms across languages and cultures that is portable and readily implementable across jurisdictions. We discuss the operational design of this multi-regional, hybrid operation and assess its strengths and limitations relative to our other cases. Furthermore we demonstrate how the exercise provided useful data for building new tools, such as testing benchmarks, and identified areas for further focus and development, which the government of Singapore is actively pursuing.

\section{Conclusions}

Ultimately, we argue that these types of exercises can empower public entities to deliver their responsible AI obligations and, in so doing, meaningfully advance the ethical development, deployment, and governance of AI. First and foremost, they draw upon the diversity of an entire national jurisdiction (or entire regions) to supply a diverse sample of capable red teamers that provide authentic grounding in the norms, values, and discourses that construct different deployment contexts. Second, by partnering with a public entity, this type of exercise more meaningfully serves the public interest and indeed the global public interest, ensuring “civil society retains its agenda-setting power over the development and deployment of AI” [12]. Third, by involving AI application developers, this type of exercise provides meaningful red teaming support to Small and Medium-size Enterprises (SME) that otherwise may not be able to meet their responsible AI obligations. Finally, we believe that these exercises provide validation of a scalable model that most AI-developing states or regions can employ to ensure that AI systems developed for their communities are safe and beneficial. 

\section*{Acknowledgments}

We wish to thank all those who participated in each of the events discussed, specifically the CAMLIS red team, our colleagues at NIST, and Lee Wan Sie, Vanessa Wilfred, Darshini Ramiah, Clarissa Koh, and Michelle Yap at IMDA, and other participants and partners who have chosen to remain anonymous.


\begin{thebibliography}{00}
\bibitem{b1} Bullwinkel, B., Amanda Minnich, Shiven Chawla, Gary Lopez, Martin Pouliot, Whitney Maxwell, Joris de Gruyter, Katherine Pratt, Saphir Qi, Nina Chikanov, Roman Lutz, Raja Sekhar Rao Dheekonda, Bolor-Erdene Jagdagdorj, Eugenia Kim, Justin Song, Keegan Hines, Daniel Jones, Giorgio Severi, Richard Lundeen, Sam Vaughan, Victoria Westerhoff, Pete Bryan, Ram Shankar Siva Kumar, Yonatan Zunger, Chang Kawaguchi, Mark Russinovich. 2025. Lessons From Red Teaming 100 Generative AI Products. https://doi.org/10.48550/arXiv.2501.07238 [cs.AI].
\bibitem{b2} Butler, J., Vorvoreanu, M., Janßen, R., Sellen, A., Immorlica, N., Hecht, B., Teevan. J. (Eds.). 2024. Microsoft New Future of Work Report 2024. Microsoft Research Tech Report MSR-TR-2024-56. https://aka.ms/nfw2024.
\bibitem{b3} Feffer, Michael; Anusha Sinha, Wesley Hanwen Deng, Zachary C. Lipton, Hoda Heidari. 2024. Red-Teaming for Generative AI: Silver Bullet or Security Theater? arXiv:2401.15897v3 [cs.CY].
\bibitem{b4} Gillespie, T., Shaw, R., Gray, M. L., and Suh, J. (2024). AI Red-Teaming is a Sociotechnical System. Now What? arXiv:2412.09751v1 [cs.CY].
\bibitem{b5} Kennedy, Wm. Matthew and Daniel Vargas Campos. 2025. Vernacularizing Taxonomies of Harm is Essential for Operationalizing Holistic AI Safety. Proceedings of the 2024 AAAI/ACM Conference on AI, Ethics, and Society. AAAI Press, 698–710.
\bibitem{b6} Kumar, Ram Shankar Siva. 2023. Microsoft AI Red Team building future of safer AI. https://www.microsoft.com/en-us/security/blog/2023/08/07/microsoft-ai-red-team-building-future-of-safer-ai/
\bibitem{b7} Longpre, S., Kapoor, S., Klyman, K., et al. (2024). A Safe Harbor for AI Evaluation and Red Teaming. arXiv:2403.04893v1 [cs.AI].
\bibitem{b8} National Institute of Standards and Technology. 2023. Artificial Intelligence Risk Management Framework. NIST-AI-100-1. https://nvlpubs.nist.gov/nistpubs/ai/NIST.AI.100-1.pdf 
\bibitem{b9} National Institute of Standards and Technology. 2024. Artificial Intelligence Risk Management Framework: Generative Artificial Intelligence Profile. NIST-AI-600-1. https://nvlpubs.nist.gov/nistpubs/ai/NIST.AI.600-1.pdf
\bibitem{b10} Schwartz, R., Rumman Chowdhury, Akash Kundu, Heather Frase, Marzieh Fadaee, Tom David, Gabriella Waters, Afaf Taik, Morgan Briggs, Patrick Hall, Shomik Jain, Kyra Yee, Spencer Thomas, Sundeep Bhandari, Paul Duncan, Andrew Thompson, Maya Carlyle, Qinghua Lu, Matthew Holmes, Theodora Skeadas. 2025. Reality Check: A New Evaluation Ecosystem Is Necessary to Understand AI's Real World Effects. arXiv:2505.18893v4 [cs.CY].
\bibitem{b11} Singh, R., Borhane Blili-Hamelin, Carol Anderson, Emnet Tafesse, Briana Vecchione, Beth Duckles, and Jacob Metcalf. 2025. Red Teaming in the Public Interest. Data and Society Research Institute and AI Risk and Vulnerability Alliance.
\bibitem{b12} United Nations Educational, Scientific, and Cultural Organization (UNESCO). 2022. Recommendation on the Ethics of AI. https://unesdoc.unesco.org/ark:/48223/pf0000381137
\bibitem{b13} Weidinger, L.; Rauh, M.; Marchal, N.; Manzini, A.; Hendricks, L. A.; Mateos-Garcia, J.; Bergman, S.; Kay, J.; Griffin, C.; Bariach, B.; Gabriel, I; Rieser, V.; Isaac, W. 2023. Sociotechnical Safety Evaluation of Generative AI Systems. https://doi.org/10.48550/arXiv.2310.11986 [cs.AI].
\end{thebibliography}
\end{document}